# Condensation on Slippery Asymmetric Bumps

Kyoo-Chul Park[1,2], Philseok Kim[2,3], Neil He[4], Joanna Aizenberg[1,2,5]

[1] School of Engineering and Applied Sciences, Harvard University, Cambridge, MA, USA
[2] Wyss Institute for Biologically Inspired Engineering, Harvard University, Cambridge, MA, USA
[3] SLIPS Technologies, Inc. Cambridge, MA, USA
[4] Department of Materials Science and Engineering, Case Western University, Cleveland, OH, USA
[5] Department of Chemistry and Chemical Biology, Harvard University, Cambridge, MA, USA

**Abstract**

Bumps are omnipresent from human skin to the geological structures on planets, which offer distinct advantages in numerous phenomena including structural color, drag reduction, and extreme wettability. Although the topographical parameters of bumps such as radius of curvature of convex regions significantly influence various phenomena such as anti-reflective structures and contact time of impacting droplets, the effect of the detailed convex topography on growth and transport of condensates have not been clearly understood. Inspired by the millimetric bumps of the Namib Desert beetle, here we report the identified role of radius of curvature and width of bumps with homogeneous surface wettability in growth rate, coalescence and transport of water droplets. Further rational design of asymmetric topography and synergetic combination with slippery coating simultaneously enable self-transport, leading to unseen five-fold higher growth rate and an order of magnitude faster shedding time of droplets compared to superhydrophobic surfaces. We envision that our fundamental understanding and innovative design of bumps can be applied to lead enhanced performance in various phase change applications including water harvesting.

The growth and transport dynamics of condensates on topographically heterogeneous surfaces is of the essence for a wide range of ubiquitous phenomena and applications from condensed water on a cold beverage bottle, to fogging on glasses and windshields[1], prey-catching means of biological systems[2], frost weathering[3], self-assembly[4], chemical vapor deposition[5], self-cleaning by jumping droplets[6], water harvesting[7,8], latent heat transfer[9-12], to name a few. To enhance the overall phase change efficiency required for numerous applications, not only to facilitate the growth of condensates but also to easily transport them to a desired direction are significantly important for making fresh sites for renucleation and regrowth[9,10,12,13].

Although intensive efforts including studies inspired by the bump structure on Namib Desert beetles ($R_{bump}$ ~ 1 mm, Fig. 1a) have been invested for the past decades, the role of convex bump topography in the dynamics of growing and moving condensates is still unclear and controvertial[6,7,14-21]. On the one hand, at microscale, convex topography is disadvantageous when the radius of curvature is smaller than five-fold of the critical radius of nuclei (*i.e.*, concave regions show preferential condensation)[22]. On the other hand, some studies tested the effect of bumps on the preferential condensation at micrometric length scale; however, similar to results observed on the flat surfaces with homogeneous wettability, the microtexture showed random nucleation without heterogeneous wettability that plays a major role in preferential condensation[19,20]. In addition, the detailed role of topographical parameters (*e.g.*, radius of curvature, width, *etc*.) in preferential growth and transport has been mostly unexplored and still not clearly understood[15,19,20,23]. Furthermore, the current designs of heterogeneous wettability-based preferential condensation are not efficient to transport the condensates due to the strong pinning points created by the high contact angle hysteresis of highly wettable region[19,20].

To facilitate the transport of condensates, superhydrophobic surfaces (SHS), which make use of an entrapped air layer to reduce the friction at the condensate-solid surface, have been considered as the most promising technique for fast removal of condensates[6,9,16]. These surfaces are, however, still plagued with inevitable problems: the air layer inhibits heat transfer for phase change due to the low thermal conductivity[9]; fully wetted droplets produce highly pinned condensates[9]; the surfaces get easily contaminated with both organic and inorganic particles[11,24]; and cannot self-heal[11,24]. To circumvent the aforementioned limitations of SHS, the lubricant on slippery liquid-infused porous surfaces (SLIPS) created by proper choice of a lubricant and nanotextured surfaces can be utilized. In contrast to SHS, SLIPS has been shown to exhibit negligible contact angle hysteresis of various condensates, excellent thermal contact due to the absence of air layer, extreme droplet mobility and shedding, particularly over a broad range of temperature and humidity conditions[10,11,13,24,25]. SLIPS have been recently shown to induce dropwise condensation with the smallest droplet volume that leads to high condensation efficiency[10,13]. The SLIPS design provides immediate self-repair by wicking into damaged sites in the underlying substrate; and can be chosen to repel any immiscible fluid[24,26]. However, due to the homogeneous chemical affinity to water, any significant effect of bump topography on the preferential condensation or transport of condensates has not been reported yet[10,13].

Here, we report topographically preferential condensation followed by guided coalescence and self-transport of water droplets on slippery asymmetric bumps. We demonstrate that convexity of bump topography (Fig. 1b) at mesoscale can solely enhance localized diffusion flux (Fig. 1c) within the depletion layer and adjacent convex curvature regions can lead to faster droplet growth by systematic understanding the effect of topographical parameters (Fig. 1d). The experimental analysis on the early and late stages of droplet growth confirms the effect of focused diffusion flux and the role of topographical parameters in growth rate, coalescence and transport, respectively. Further we introduce the rational design of asymmetric topography and synergetic combination with SLIPS (Fig. 1e) for the optimal strategy to guide coalescence and self-transport condensed droplets, yielding order of magnitude faster droplet shedding (Fig. 1f).

The central concept of preferential condensation on the apex of spherical-cap-shaped bumps at mesoscale is based on the focusing effect of diffusion flux on to the convex topography, within the depletion layer $\delta$ (Fig. 2a, b). Assuming no convective flow within the boundary layer, superposition of diffusion fluxes on a spherical cap and flat surrounding (analogous to electric field[27]) shows that the magnitude of diffusion flux $J \sim F(\varphi)$ (see Supplementary Information (SI) for derivation) on the apex of hemispherical bumps ($\varphi = \pi/2$) is approximately 1.4 times greater compared to a flat surface. As shown in Fig. 2b, the intensity of focused diffusion flux increases as the spherical cap angle ($\varphi$) increases because the flux is independent of radius of curvature ($\kappa^{-1}$). Rigorous numerical calculation (COMSOL-Multiphysics) using axisymmetry of hemisphere confirms this enhanced diffusion flux of ~1.6 times. Fig. 2c shows the effect of focused diffusion flux on the initial growth rate and gradual difference in the diameter of water droplets. It should be noted that both numerical calculation and experiment results on spatial preferential condensation were obtained that the depletion layer thickness is more than ten times greater than the height of the bumps[23].

To quantitatively investigate the droplet growth dynamics on spherical-cap-shaped bumps within the same depletion layer thickness condition, bumps with the same height ($H$ = 1mm << $\delta$ = $10^1$ mm)[7] and spacing ratio ($P_{pattern}/R_{bump}$ =3) have been used. Note that as spherical cap angle ($\varphi$) increases, the radius of curvature ($\kappa^{-1}$) of the bump decreases under the constraint of same height. The numerical calculation shown in Fig. 3a predicts that decreasing radius of curvature leads to a higher magnitude of diffusion flux at mesoscale. The quantitative analysis on the droplet growth dynamics based on optical images (Fig. 3a) and the log-log plot (Fig. 3b) confirms that the actual growth of droplet diameter ($2r_{max}$) follows a power law model ($2r_{max} \sim \alpha t^\beta$). The intercept in the abscissa ($\alpha$) is greater as the radius of curvature decreases because the higher magnitude of diffusion flux leads to a higher growth rate of droplets at the early stage of condensation, resulting in the greater initial droplet size captured in the quantitative analysis as discussed in Fig. 2c. However, droplets further grow with a similar slope value ($\beta \sim 0.9$), regardless of the radius of curvature, on the bumped surfaces for the time range from $5\times10^2$ seconds to $5\times10^3$ seconds. Droplets on the bumps with a smaller radius of curvature finally reach the similar diameter to the droplets on other surfaces, leading to gradually decreasing slope in the log-log scaled plot, whilst drops on the bumps with a greater radius of curvature show a lower intercept without decreasing slope until $t \sim 5\times10^3$ sec.

As shown in the numerical calculation result of the smallest radius of curvature structure in Fig. 3a (right), for the same height, decreasing radius of curvature leads to a higher diffusion flux. However, a smaller radius of curvature does not rapidly produce droplets that are great enough to shed because of the decreasing slope of growth rate in the late stage captured in Figure 3b. To find a way to overcome this tradeoff, we investigated rectangular bumps that have the same radius of curvature but more area for droplets to grow further. On rectangular bumps, not only the corner region with small radius of curvature shows the maximum rate of flux but also the region on the top flat region shows enhanced diffusion flux as shown in the numerical calculation (Fig. 3c). The level of diffusion flux on the corners and top flat regions turned out to be influenced by the radius of curvature of the corner and width, respectively. The enhanced flux effect is observed when the edges are close each other. Moreover, we observed the forced coalescence during the droplet growth after which the droplet size is comparable to the width of edge, as captured by the dotted line in Fig. 3d. Those synergetic droplet growth mechanisms by both (i) the superposed focusing diffusion flux created by adjacent corner regions with a small radius of curvature and (ii) the forced coalescence could lead faster droplet growth even compared to spherical-cap-shaped bumps with the same height.

Exploiting the fundamental understanding of focusing effect of diffusion flux around convex topography ($\kappa^{-1} > 0$) and forced coalescence by topographical constraint ($W$), we rationally engineered an optimal mesoscale surface topography and coating to maximize the growth rate of condensed droplets and to minimize the time for initiating droplet shedding at the same time. First, we started with the rectangular bumps with a narrow width ($W = 0.4$ mm), the minimum value determined by the thickness of the aluminum sheet used for the high throughput fabrication method. We then replaced the sudden change of height at the edge of rectangular bumps in the direction of shedding with a tangential connection to the flat region (by modifying rectangular bumps with the tangential fillet) because the shedding droplets on the bumps are highly pinned on the edge due to the mesoscale concave curvature. To prevent a rapidly-growing droplet from running off the raised region of bumps and pinning on the side of rectangular inclined structure when the droplet size is much greater than the width of the bumps, we designed gradually increasing width in the direction of shedding. The topographical modification results in an asymmetric bump structure elongated from the rectangular bump, starting with raised a small width ($W = 0.4$ mm) and ending with a wider width tangentially connected to the flat region (Fig. 4a), which is similar to the asymmetric bumps pointing outward, found on the back of Namib Desert beetles (Fig. 1a).

In addition to the topographical pinning effects created by mesoscale concave corners, another pinning primarily due to the contact angle hysteresis also inhibits fast shedding of condensed droplets[24]. When the resultant asymmetric bump topography is combined with SLIPS (Fig. 4b) that have negligible friction with droplets (SLIPS-A hereafter), not only the SLIPS readily move droplets but also the gradually increasing width of the asymmetric bump pushes droplets toward the wider width, even against gravity as shown in Fig. 4c. This interesting phenomenon can be understood as the curved boundary of asymmetric bumps makes a guide to the lowest energy state where the diameter of droplet is equal to the width of flat region of asymmetric bumps (Fig. 4d). This motion induced by the combination of asymmetry of the designed bumps and SLIPS can further facilitate coalescence of small droplets on the path of shedding and promote fast shedding of condensed water droplets by gravity. As a droplet grows faster by focused diffusion flux and initial forced coalescence on the apex of the unique slippery asymmetric bump, the droplet is guided to move to the wider region by the resultant interaction among capillary force by the asymmetry of the bump, friction force of SLIPS, and gravitational force. The droplet further grows by the guided coalescence with the small droplet on the path and once a droplet exits from the convex asymmetric topography and reaches the tangential conjunction point, the droplet grows large enough to continue the journey by gravitational force overcoming small friction of underlying SLIPS and then sheds much earlier even compared to the small droplets on the flat region as shown in Fig. 4e.

The synergetic combination between the rationally designed asymmetric bumps and SLIPS, was confirmed by the experimental results that fast growth and shedding of droplets on this engineered surface take place even in a high supersaturation condition at which SHS have shown to fail and perform worse than untreated surfaces[9] (see Methods for detailed experimental condition). These surface features, together with nearly friction-free SLIPS, dramatically reduce the time required for a growing droplet to reach the smallest size (or volume) for spontaneous shedding. In the quantitative analysis shown in Fig. 4e, droplets on SLIPS-A and SHS-A (*i.e.,* SHS with the asymmetric bumps) in the early stage are greater than those on SLIPS and SHS due to the fast droplet growth effect by focusing of diffusion flux. In particular, during the self-transport on the asymmetric bumps represented by red shade, SLIPS-A shows the highest slope ($\beta$ = 5.0) which is more than five-fold of the maximum slope value predicted by droplet growth dynamics theory[7] and observed in typical droplet growth dynamics. As a result, the asymmetric bumps with SLIPS showed order of magnitude faster production of ~3 times smaller size of first shedding droplets, compared to SHS.

In summary, we have achieved fast droplet growth by designing optimal topography and surface wettability based on systematic understanding effect of topographical components such as radius of curvature and width of convex topography within depletion thickness. By employing further fast droplet growth by self-transport on asymmetric bump, we demonstrated fast droplet growth rate and production of first shedding droplets that break through the theoretical limit on flat surfaces. We envision that this faster droplet growth and transport enabled by both rational design of the asymmetric topography and synergetic combination with the slippery coating can be further applied to applications including efficient water condensation.

**Acknowledgements**

This research was supported by the Department of Energy/ARPA-E award #DE-AR0000326. Part of this work was performed at the Center for Nanoscale Systems (CNS) at Harvard University, supported under ARPA-E award #DE-AR0000326. NH would like to thank National Science Foundation Research Experiences for Undergraduates (REU program), co-funded by Wyss Institute.

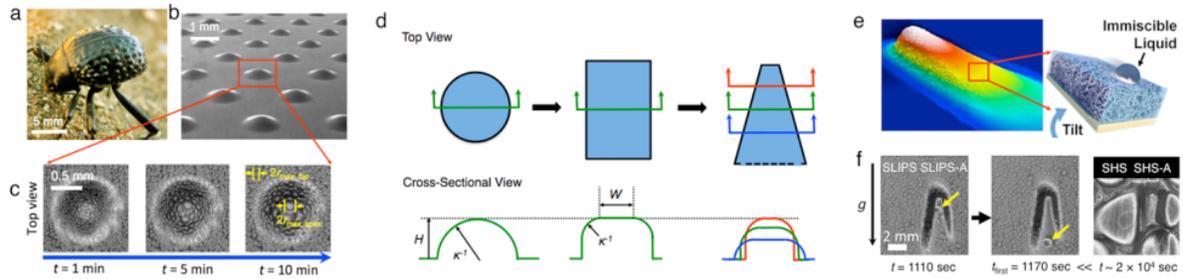

**Figure 1**. An overview of the Namib Desert beetle-inspired topographically preferential growth and shedding of droplets on various bumps. (a) A Namib Desert beetle (*Stenocara gracilipes*) with asymmetric bumps. (b) A scanning electron microscope (SEM) image of artificial spherical-cap-shaped bumps. (c) Time-lapsed images of the water droplet growing by condensation on the hydrophobic spherical-cap-shaped bumps. The diameter of droplets near the convex apex is the greatest compared to the surrounding concave and flat regions. (d) Top and cross-sectional views of spherical-cap-shaped, rectangular, and asymmetric bumps with essential topographical parameters for growth and shedding of droplets. The dotted line on the bottom of the asymmetric bump represents a tangential connection (*e.g.*, fillet) from the raised top region of the bump to the surrounding flat surface. (e) An optical profilometer image of an asymmetric slippery bump with a schematic illustration of slippery liquid-infused porous surfaces (SLIPS) coating. (f) Time-lapsed images of droplets condensed on SLIPS and SLIPS with an asymmetric bump (SLIPS-A), compared to the superhydrophobic surfaces (SHS) and SHS with the same size asymmetric bump (SHS-A). The time for producing the first shedding droplet ($t_{first}$) on SLIPS-A is order of magnitude faster than that on SHS or SHS-A.

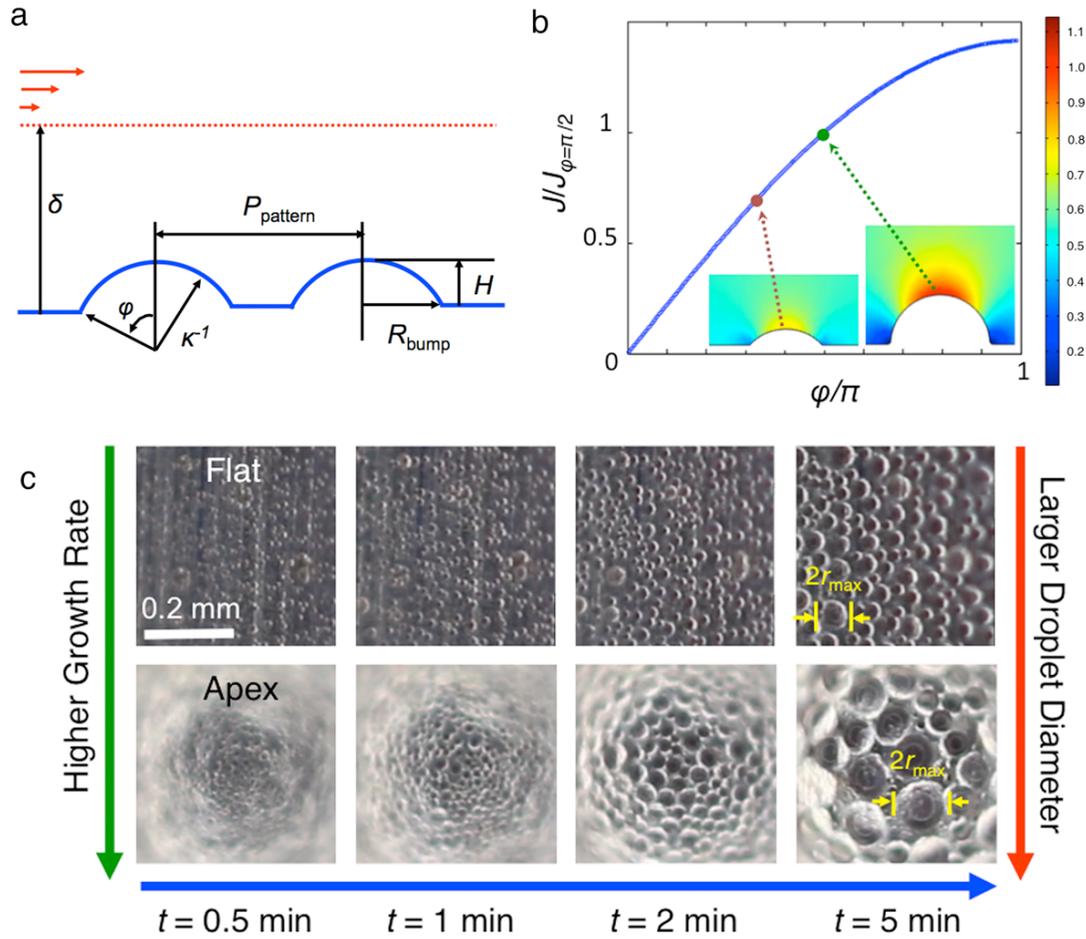

**Figure 2**. Calculated focusing effect of magnitude of diffusion flux and resultant topographically preferential condensation on the convex apex of spherical-cap-shaped bumps assuming $\delta \gg H$ (*i.e.*, without the constraint of the same $H$). (a) A schematic illustration of cross-sectional view of spherical-cap-shaped bumped surfaces (blue solid line) with topographical parameters. The dotted horizontal red line represents the location of the plane below which diffusion is the dominant mechanism of mass transport (*i.e.*, depletion layer). Above the dotted line, convection cannot be ignored as indicated by the red arrows. (b) The maximum magnitude of diffusion flux ($J$) on the apex of spherical-cap-shaped bumps normalized by the value of a hemispherical bump ($J_{\varphi=\pi/2}$) with the same radius of curvature $\kappa^{-1}$. The two insets represent magnitude of diffusion flux calculated by COMSOL-Multiphysics. The color bar is also normalized by $J_{\varphi=\pi/2}$. (c) Optical images (top view) of condensed droplets on both flat area and the apex of a bump with the homogenous wettability (*i.e.*, hydrophobic, $\theta = 113 \pm 24°$) in the same sample. At the early stage of condensation (left), the growth rate of the droplets on the apex of the bump is higher than that of the flat surface, resulting in the greater droplet size after subsequent steps of condensation and coalescence (bottom).

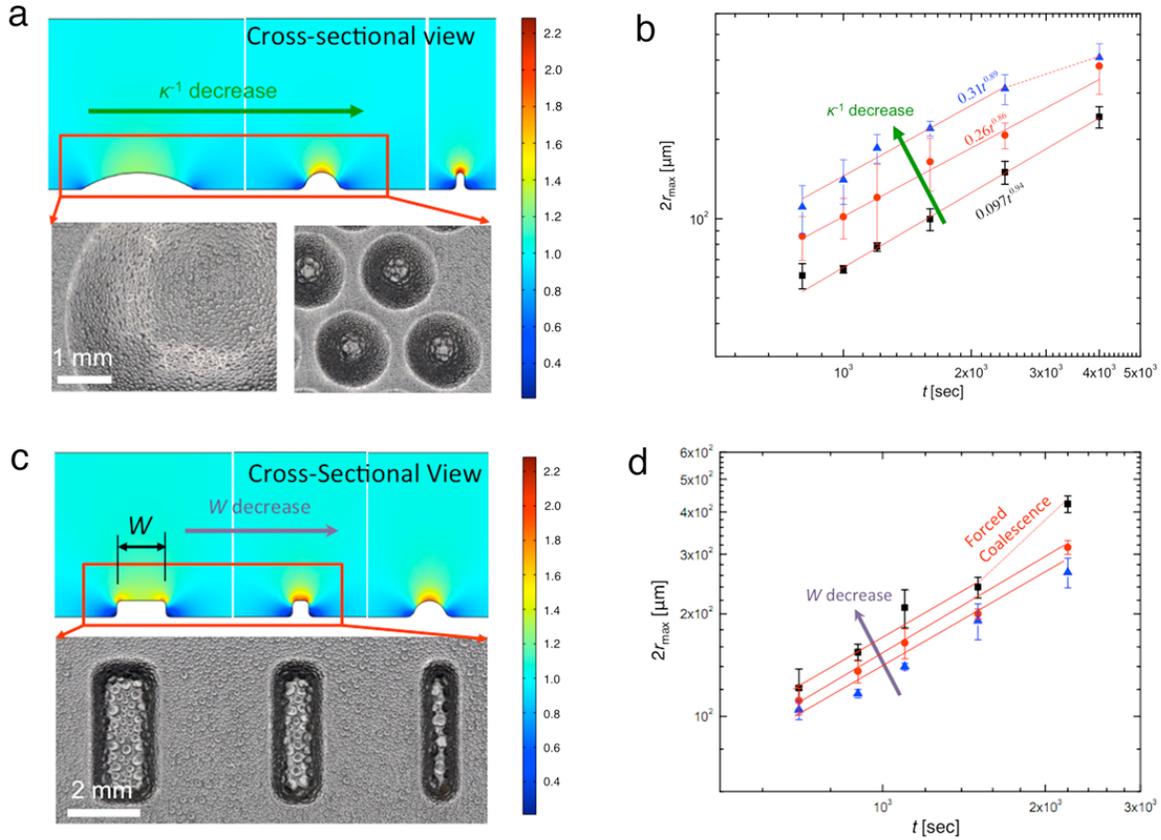

**Figure 3**. Effects of topographical parameters on droplet growth dynamics on top of the apex of various bumps ($\theta$ = 113 ± 24°) with the constraint of the same $H$. (a) As the radius of curvature $\kappa^{-1}$ decreases (or the spherical cap angle ($\varphi$) decreases), the calculated magnitude of diffusion flux represented by color increases, which is confirmed by the optical images of condensed droplets on two representative spherical-cap-shaped bumps. (b) For three representative spherical-cap-shaped bumps, the y-intercept values are different whilst the slopes of the droplet growth dynamics are approximately similar. (c, d) Effect of width of rectangular bumps on the magnitude of diffusion flux and forced coalescence. The smaller width ($W$) increases the focusing effect of diffusion flux created by the convex regions with small radius of curvature, leading to a higher magnitude of diffusion flux value compared to the hemispherical bump with the same height. Once the droplets on the edge where convex curvature exists reach the smaller width of the top flat area of rectangular bumps, forced coalescence occurs, leading to the higher slope in (d), which cannot be explained by the droplet growth dynamics model based on condensation and coalescence on flat surfaces. The color bars are normalized by the magnitude of diffusion flux on flat surfaces.

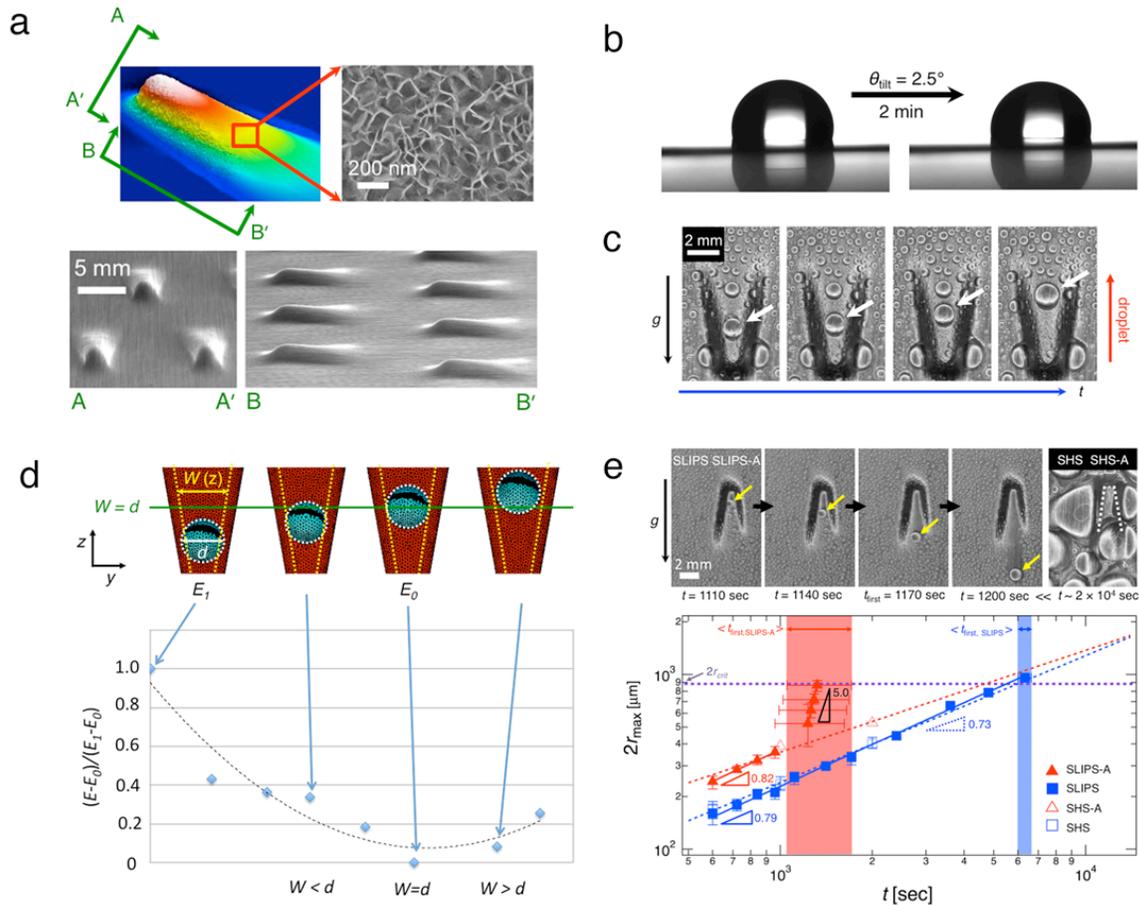

**Figure 4**. Self-transport and guided coalescence on asymmetric slippery bumps. (a) An optical profilometer image with SEM images of asymmetric bumps with nanostructure (*i.e.*, boehmite) used in the experiments (oblique views). (b) Goniometer images of a water droplet, showing water contact angle ($\theta_{SLIPS} = 101 \pm 2°$) and slippery property of SLIPS. (c) Time-lapsed optical images of condensed water droplets on an asymmetric bump. The droplet pointed by the white arrow climbs up against gravity with the tilting angle of 90°. This synergetic effect is observed only when an asymmetric bump is combined with SLIPS. (d) Energy landscape normalized by the difference between the maximum and minimum value of Gibbs free energy of the asymmetric bump-droplet-vapor system. The values were obtained by using finite element method based numerical calculation software, Surface Evolver, using the same volume of the water droplet on SLIPS-A. The droplet shows the minimum energy state when $W = d$. This calculation explains the combined effect of the radius of curvature and width of the asymmetric bump on the self-transport of condensed droplet toward the gradually wider width. In an actual condensation situation, droplet size also increases by subsequent guided coalescence while the droplet moves, which creates favorable effects. (e) Growth dynamics and the first shedding time ($t_{first}$) of the maximum sized water droplets on the four different surfaces when the direction of droplet motion is aligned with gravity. Both SLIPS-A and SHS-A ($\theta_{SHS} = 155 \pm 2°$) show faster localized droplet growth in the early stage ($t < 10^3$ sec) compared to flat surface cases. SLIPS-A exhibits the unseen fast growth denoted by black triangle (slope of approximately five times greater than slope values in typical droplet growth dynamics) and shortest $t_{first}$ due to the additional driving force of self-transport on the asymmetric bump with the very small contact angle hysteresis of condensates on SLIPS although the critical shedding droplet size (denoted by horizontal purple dotted line) is the same as SLIPS. Both SHS-A and SHS did not produce shedding droplets until $t \sim 18{,}000$ sec.